\documentclass[lettersize,journal]{IEEEtran}

\usepackage{color}
\usepackage{tabularray}

\usepackage{amsmath,amsfonts}
\usepackage{amssymb}
\usepackage{algorithmic}
\usepackage{array}
\usepackage[caption=false,font=normalsize,labelfont=sf,textfont=sf]{subfig}
\usepackage{textcomp}
\usepackage{stfloats}
\usepackage{verbatim}
\usepackage{graphicx}
\usepackage{mathrsfs}
\usepackage{CJK}
\usepackage{indentfirst}
\usepackage{cases}
\usepackage{array}
\usepackage{titlesec}
\usepackage{xcolor}

\def\BibTeX{{\rm B\kern-.05em{\sc i\kern-.025em b}\kern-.08em
    T\kern-.1667em\lower.7ex\hbox{E}\kern-.125emX}}
   
\usepackage{balance}

\begin{document}

\title{Physics-Guided Graph Neural Networks for  Real-time AC/DC Power Flow Analysis}
\author{Mei Yang, Gao Qiu, Yong Wu, Junyong Liu, \emph{Member, IEEE}, Nina Dai, Yue Shui, Kai Liu, \emph{Senior Member, IEEE}, Lijie Ding
\vspace{-4ex}
\thanks{This work was supported in part by the Fundamental Research Funds for the Central Universities (YJ2021162), in part by Chongqing Natural Science Found Project (cstc2022NSCQ-MSX4086), and in part by the Science and Technology Department of Sichuan Province (2021LDTD0016). (Corresponding author: \emph{Gao Qiu})

Mei Yang, Gao Qiu, Junyong Liu and Kai Liu are with the College of Electrical Engineering, Sichuan University, Chengdu 610065, China (email: 20140046@sanxiau.edu.cn; qiugscu@scu.edu.cn; liujy@scu.edu.cn; kailiu@scu.edu.cn)

Mei Yang and Nina Dai are with the Key Laboratory of Information and Signal Processing, Chongqing Three Gorges University, Chongqing 404020, China. (email: 20140046@sanxiau.edu.cn; 20040001@sanxiau.edu.cn)

Yong Wu is with Megvii Inc, Chengdu 610039, China. (email: wuyong@megvii.com

Yue Shui is with Skill Training Center of Sichuan Electric Power Corporation of State Grid, Chengdu 610000, China. (email: 178332804@qq.com))

Lijie Ding is with State Grid Sichuan Electric Power Research Institute, Chengdu 610000, China (email: ding-lijie@163.com)
}}

\markboth{Journal of \LaTeX\ Class Files,~Vol.~18, No.~9, September~2020}%
{How to Use the IEEEtran \LaTeX \ Templates}
\maketitle
\begin{abstract}
The increasing scale of alternating current and direct current (AC/DC) hybrid systems necessitates a faster power flow analysis tool than ever. This letter thus proposes a specific physics-guided graph neural network (PG-GNN). \textcolor{black} {The tailored graph modelling of AC and DC grids is firstly advanced to enhance the topology adaptability of the PG-GNN}. \textcolor{black} {To eschew unreliable experience emulation from data, AC/DC physics are embedded in the PG-GNN using duality.} \textcolor{black} {Augmented Lagrangian method-based learning scheme is then presented to help the PG-GNN better learn nonconvex patterns in an unsupervised label-free manner. Multi-PG-GNN is finally conducted to master varied DC control modes. Case study shows that, relative to the other 7 data-driven rivals, only the proposed method matches the performance of the model-based benchmark, also beats it in computational efficiency beyond 10 times.} 
\end{abstract}
\vspace{-2ex}
\section{Introduction}
\IEEEPARstart{H}{igher} \textcolor{black}{power transmission efficiency of AC/DC hybrid systems has widely benefited renewable energy consumption \cite{r1}}. Faster AC/DC power flow analysis is thus necessary more than ever, due to significant uncertainties from renewable energy, massive scale of interconnected grids, and nonlinearity of DC lines and converters \cite{r2}. This is the key for real-time dispatch and control in AC/DC hybrid systems.

Rich model-based methods are available for AC/DC power flow, spanning from the basic unified methods \cite{r3} and sequential methods \cite{r3a}, to the their enhanced parallel versions \cite{r2, r4}. However, issues, such as \textcolor{black}{slow convergence rate and diverse control modes}, can still beget heavy computing burden of power flow analysis in large-scale AC/DC hybrid systems. Emerging data-driven methods provide faster solutions \cite{r5, r6, r7}, but their generalizability can be limited by the absence of \textcolor{black}{physics embedding and topology patterns}. 

\textcolor{black}{Graph neural network (GNN) may be the prospective fix for above concerns, only when it is carefully trained to follow physics. Two alternatives exist towards this end, one is to emulate systematic process of power flow solvers, such as Newton-Raphson (NR) solver \cite{r11}, Hades2 solver \cite{r9}, etc. However, this method may be slow-paced when DC control mode switches, since the solving workflow must be revisited to hit new control mode. Treating power flow equations directly as loss function is the other candidate \cite{r10}. This idea is better, since a one-time feedforward computation is all we need to catch a power flow solution, provided that GNN has been well-trained. The remaining issue is how to embed all AC/DC physical principles into loss function. Penalizing them is pragmatic but imprudent, as it is formidable to handpick penalty weights from scratch. Besides, DC control mode switching has not been studied by the above methods.} 

To fill the above gaps, an augmented Lagrangian method (ALM)-based physics-guided GNN (PG-GNN) is proposed. \textcolor{black}{On the tailored graph modelling of AC and DC grids, AC/DC operational physics are analytically embedded into GNN learning process}. Then, the learning task is recast as a parameterized duality problem of AC/DC power flow model. \textcolor{black}{To yield tolerable performance of the PG-GNN against nonconvex physics, the ALM-based gradient decent algorithm is proposed to cope with the above problem. At last, a multi-PG-GNN decision framework is presented, with a view to enable DC control mode switching.} 

\vspace{-0.8em}
\section{Power Flow Model for AC/DC Hybrid Systems}
In this letter, we apply the quasi-steady state model of AC/DC hybrid system. Let $\mathcal{N}_{PV}$, $\mathcal{N}_{ac}$, $\mathcal{N}_{dc}$, and $\mathcal{N}_{pcc}$ stands for PV bus set, AC bus set, DC bus set, and the set of point of common coupling, respectively. $\mathcal{N}_{pcc}$ implies the bus connected by both DC links and AC branches. Following the definitions, the AC/DC power flow is modelled as follows ($\forall i  \in   \mathcal{N}_{ac},  j  \in  \mathcal{N}_{ac} \cup  \mathcal{N}_{pcc}, k  \in   \mathcal{N}_{dc}, p \in  \mathcal{N}_{PV}, c \in  \mathcal{N}_{pcc}$):

\begin{subequations}     \label{eq_1}
\begin{equation} 
\label{eq_1a}
P_{i}^{inj} - \sum_{j\in i} P_{ij}(V_{i},V_{j},\delta_{ij}) = 0
\end{equation}
\begin{equation}   
\label{eq_1b}
Q_{i}^{inj} - \sum_{j\in i} Q_{ij}(V_{i},V_{j},\delta_{ij}) = 0
\end{equation}
\end{subequations}
\begin{subequations}     
\label{eq_2}
\begin{equation} 
\label{eq_2a}
P_{c}^{inj} - \sum_{j\in c} P_{cj}(V_{c},V_{j},\delta_{ij}) \pm  V_{d,k}^{re/iv}I_{d,k}  = 0
\end{equation}
\begin{equation}   
\label{eq_2b}
 Q_{c}^{inj} \sum_{j\in c} Q_{cj}(V_{c},V_{j},\delta_{ij}) V_{d, k}^{re/iv}   I_{d,k}   \tan  \varphi  _{k} ^{re/iv} = 0
\end{equation}
\end{subequations}
\begin{equation}
\label{deqn_ex3}
V_{p} = V_{p}^{ref}
\end{equation}
\begin{subequations}
    \label{eq_4}
    \begin{equation} 
    \label{eq_4a}
    V_{d,k}^{re} =  {3\sqrt{2}}K_{ck}^{re} V_{c} \cos\alpha_{k} / {\pi}  -  {3}X_{ck}^{re} I_{d,k} / {\pi}
    \end{equation}
    \begin{equation} 
    \label{eq_4b}
    V_{d,k}^{iv} =  {3\sqrt{2}}K_{ck}^{iv} V_{c} \cos\gamma_{k} / {\pi}  -  {3}X_{ck}^{iv} I_{d,k} / {\pi}
    \end{equation}
\end{subequations}

\begin{equation}
\label{eq_5}   
R_{dc}I_{d,k}=V_{d,k}^{re} - V_{d,k}^{iv}
\end{equation}
\begin{equation} 
\label{eq_6}
{\cos\varphi_{k} ^{re/iv} = {V_{d,k}^{re/iv}}  / ({{3\sqrt{2}}K_{ck}^{re/iv} V_{c}} / {\pi})}
\end{equation}
\begin{subequations}     
\label{eq_7}
\begin{equation} 
\label{eq_7a}
 \text{Control model 1: } {I_{d,k}^{re}\!-\!I_{d,k}^{re,ref}\! = \!0},  {V_{d,k}^{iv}\!-\!V_{d,k}^{iv,ref} = 0}
\end{equation}
\begin{equation}  
\label{eq_7b}
\text{Control model 2: } 
 {\alpha_{k} - {\alpha}_{k}^{min} =  0,I_{d, k}^{iv}  -   I _{d, k}^{iv,ref}  =  0 }
\end{equation}
\end{subequations}
where $ \textit{K}_{ck}^{re} $ and $ \textit{K}_{ck}^{iv} $ stand for the ratio of the converter transformer at the rectifier-side and inverter-side $k$. \textcolor{black}{Note that, superscript $re/iv$ stand for rectifier or inverter, respectively.} $\textit{X}_{ck}^{re} $ and $\textit{X}_{ck}^{iv} $ represent the reactance of the  converter connected at the inverter and rectifier DC bus $k$. \textcolor{black}{ $\alpha_{k}$ is firing angle of the $k$th rectifier, and $\gamma_{k}$ is extinction angle of the $k$th inverter.} $V_{d,k}^{re / iv}$\ represent the rectifier/inverter voltage of $k$th DC line. $R_{dc}$ and $I_{d,k}$ represent resistance and current of the $k$th DC line, respectively. $V$\ indicates the nodal voltage amplitude, $j\in i$\ implies that the $j$th bus is adjacent to the $i$th bus through branch $ij$. $P_{ij}(V_{i},V_{j},\delta_{ij})$ and $Q_{ij}(V_{i},V_{j},\delta_{ij})$ are function to calculate active and reactive transmission power of line $ij$, respectively. $ \delta_{ij}$ is voltage phase angle difference between bus $i$ and bus $j$. The sign of the third term on the right hand side of ($\ref{eq_2}$) is negative if it specify the rectifier side, and is positive otherwise. ($\ref{eq_7}$) are governing equations of the DC system. Here we only provide two common control modes. The subscript ${ref}$ indicates the control reference.  

\vspace{-0.8em}
\section{The Proposed Methodology}
\subsection{Graph modeling for AC/DC power flow analysis}
A power system can be represented as directed graph $\mathcal{G} = (\mathcal{V}, \mathcal{E}, \bold{A})$, where $\mathcal{V}$, $\mathcal{E}$ and $\bold{A}$ are settled upon the set of buses, that of branches, and adjacency matrix, respectively. \textcolor{black}{GNN naturally involves these properties for well topology-adaptability. For an AC/DC hybrid system, GNN is not easy to create since features necessary to characterize AC grids, DC grids and their bridges do not always match each other. Thus, this letter proposes an alternative to unify these features. As shown in Fig. 1, by introducing converters into nodes, and DC transformers and lines into edges, a DC grid can be unified in a common graph of AC grid.} In this sense, for an AC/DC system with $N$ buses, $\textcolor{black}{b}$ branches and $\textcolor{black}{a}$ DC lines, its corresponding GNN owns $N+2\textcolor{black}{a}$ nodes and $\textcolor{black}{b}+3\textcolor{black}{a}$ edges:
\vspace{-1ex}
\label{eq_8}
\begin{equation}
  \boldsymbol{H}(\boldsymbol\cdot)^{(0)} = \boldsymbol{x} \triangleq \left [ \boldsymbol{x}^\mathcal{V}, \boldsymbol{\textcolor{black}{Z}}^{\mathcal{V} \leftarrow \mathcal{E}} \boldsymbol{x}^\mathcal{E} \right ] \vspace{-1ex} 
\end{equation}
where $\boldsymbol{x}^\mathcal{V}\! \!=\!\! [ \boldsymbol{x}^{\mathcal{V}}_1, \dots, \boldsymbol{x}^{\mathcal{V}}_i, \dots, \boldsymbol{x}^{\mathcal{V}}_{\textcolor{black}{N\!+\!2a}} ]^\mathrm{T} \! \in \! \mathbb{R}^{\textcolor{black}{(N\!+\!2a)} \! \times \! 2}$ and particularly, $ \boldsymbol{x}^{\mathcal{V}}_i \! = \! [ {P}^g_{i} \! - \! {P}^L_{i}, \! V_i ]^\mathrm{T}, i \! \in \! \mathcal{N}\!_{PV} $; \textcolor{black}{ For rectifiers, $\boldsymbol{x}_k^{\mathcal{V}} \! = \!  [\boldsymbol{\alpha}^{min},  {X}_{c k}^{re} {I}_{d,k}^{re,ref}  ]$ (mode 1) or $\boldsymbol{x}_k^{\mathcal{V}}\! = \! [\boldsymbol{\alpha}^{min}, {X}_{ck}^{re} {I}_{d,k}^{iv,ref}]$ (mode 2). For inverters, $\boldsymbol{x}_k^{\mathcal{V}} \! = \! [\boldsymbol{\gamma}^{min}, \boldsymbol{V}_{d,k}^{iv,ref} ]$ (mode 1) or $\boldsymbol{x}_k^{\mathcal{V}} \! = \! [\boldsymbol{\gamma}^{min}, {X}_{c,k}^{iv} {I}_{d,k}^{iv,ref}  ]$ (mode 2), } $k \! \in \! \mathcal{N}_{dc} $; $\boldsymbol{x}^{\mathcal{V}}_i= [{P}^g_{i}-{P}^L_{i}, {Q}^g_{i} \! - \! {Q}^L_{i}]^\mathrm{T}, i \! \in \! \mathcal{N}_{PQ} \cup \mathcal{N}_{pcc} $; $ \boldsymbol{x}^{\mathcal{V}}_i \! =  \! \ [{P}^g_{i} \! - \! {P}^L_{i}, V_{i}]^\mathrm{T}, i \in \mathcal{N}_{V \delta} $, $\mathcal{N}_{V \delta}$ denote the set of swing bus. The edge features of the input are structured by the branches, which are denoted by $\boldsymbol{x}^\mathcal{E} \! \in \! \mathbb{R}^{(\textcolor{black}{b}+3\textcolor{black}{a}) \times 2}$. Specifically, for AC branches, $\boldsymbol{x}^{\mathcal{E}}\!=\![ G_{\mathcal{E}}, B_{\mathcal{E}}] $, where $ G_{\mathcal{E}}$ and $ B_{\mathcal{E}}$ are admittance parameters of line $\mathcal{E}$. For rectifiers and  inverters, $\boldsymbol{x}^{\mathcal{E}} \! = \! [ K_{ck}^{re,min}, K_{ck}^{re,max} ]$. For DC branches, $\boldsymbol{x}^{\mathcal{E}}= [ {R}_{dc},  {I}_{d,k}^{re,ref} ]$ (mode 1) or $\boldsymbol{x}^{\mathcal{E}}= [ {R}_{dc},  {I}_{d,k}^{iv,ref}]$ (mode 2). $\boldsymbol{\textcolor{black}{Z}}^{\mathcal{V} \leftarrow \mathcal{E}} \! \in \! \mathbb{R}^{(N+2a) \times (\textcolor{black}{b}+3\textcolor{black}{a})} $ is a trainable transformation matrix that embeds edge features into nodes.

We then consider ChebNet to enhance GNN, as ChebNet can aggregate high-order neighbor information and result in better performance \cite{r8}. Prior to the depiction of ChebNet, we firstly calculate the graph Laplacian matrix $\boldsymbol{L}$ by subtracting $\boldsymbol A$ from degree matrix $\boldsymbol{D}$, i.e., $\boldsymbol{L} = \boldsymbol{D} -\boldsymbol{A}$. Then, the feedforward of the ChebNet is given by (9):
\vspace{-0.5em}
\begin{subequations}
\begin{equation}
\label{eq_9a}
  \boldsymbol{H}(\boldsymbol\cdot)^{(l+1)}=  \sigma \left(\begin{matrix} \sum_{j=0}^F \end{matrix} \boldsymbol{T}_j(\hat L) \boldsymbol{H}(\boldsymbol\cdot)^{(l)} \boldsymbol{\theta}_j \right), \textcolor{black}{\boldsymbol{H}(\boldsymbol\cdot)^{(0)}\!=\!\boldsymbol{x}}
  \end{equation}
\vspace{-3ex}
\begin{equation}
\label{eq_9b}
\mathcal{O} \! = \!\boldsymbol{H}  (\boldsymbol\cdot) ^{(M)} \! =  \! \sigma \left(\!\begin{matrix} \sum_{j=0}^F \end{matrix} \boldsymbol{T}_j(\hat L)  \boldsymbol{H}  (\boldsymbol\cdot) \! ^{( M \! -  \! 1  )} \boldsymbol{\theta}_j \! \right) \! =\!   \boldsymbol{\phi} ^{\theta}  (\boldsymbol x)
\end{equation}
\end{subequations}
where \textcolor{black}{$F$ represents the order of neighbor node, and} $\boldsymbol{H}(\boldsymbol\cdot)^{(l)}$ is the $l$th layer's response with the preceding layer's response as the input. The output of the GNN $\mathcal{O}$ is the response of the last layer $M$ and includes voltage and phase angle for AC bus\textcolor{black}{, and converter voltage and firing angle (or extinction angle) for DC bus}, i.e., $\mathcal{O} = \boldsymbol{\phi} ^{\boldsymbol{\theta}}  (\boldsymbol x) = \textcolor{black}{[\boldsymbol{V}, \boldsymbol\delta] \, \text{or} \, [\boldsymbol{V}, \boldsymbol\alpha (\text{or} \, \boldsymbol{\gamma})]} $. For brevity, we use $\boldsymbol{\phi}^{\boldsymbol{\theta}}({\boldsymbol x})$ to denote the ChebNet. $\boldsymbol{T}_k(\boldsymbol{\hat L})$ is Chebyshev polynomials, i.e., $\boldsymbol{T}_0(\boldsymbol{\hat L}) = \bold{I}$,  $\boldsymbol{T}_1(\boldsymbol{\hat L}) = \boldsymbol{\hat L}$,  $\boldsymbol{T}_{n+1}(\boldsymbol{\hat L}) = 2 \boldsymbol{\hat L} -  \boldsymbol{T}_{n}(\boldsymbol{\hat L})\boldsymbol{T}_{n-1}(\boldsymbol{\hat L})$, and $\boldsymbol{\hat L} =  {2} \boldsymbol{ L} / \zeta_{max} -\bold{I}$. $\zeta_{max}$ is the maximum eigenvalue of $\boldsymbol{L}$. 

\vspace{-2.5ex}
 \subsection{Learning the physics-guided graph neural networks}
\textcolor{black}{Upon the above graph modelling for AC/DC grids, we can train a topology-adaptive GNN for real-time power flow analysis. To foster the GNN to learn physics, a augmented Lagrangian method (ALM)-based training scheme is proposed based on the equivalent optimization form (10) of (1)-(7). Notably, to enable control mode switching, we hold violations of firing/extinction angle constraints to be minimization objective, which in fact relaxes the constraints.}
\vspace{-1ex}
\begin{equation}
\label{eq_10} 
 \min\limits_{\boldsymbol{\theta}}  f_{DC}^{Ang}(\boldsymbol{x}, \mathcal{\phi}^{\boldsymbol{\theta}}(\boldsymbol{x})) \ \ \ \ \rm \bold s. \rm \bold t.  \ 
\begin{cases}
  \vspace{-0.3em}
  f_{PQV}(\boldsymbol{x}, \mathcal{\phi}^{\boldsymbol{\theta}}(\boldsymbol{x})) = 0, \\
  \vspace{-0.3em}
  f_{DC}^{eq}(\boldsymbol{x}, \mathcal{\phi}^{\boldsymbol{\theta}}(\boldsymbol{x})) = 0, \\
  \vspace{-0.3em}
  f_{DC}^{Con}(\boldsymbol{x}, \mathcal{\phi}^{\boldsymbol{\theta}}(\boldsymbol{x}))  = 0, \\ 
  \vspace{-0.3em}
  f_{DC}^{K}(\boldsymbol{x}, \mathcal{\phi}^{\boldsymbol{\theta}}(\boldsymbol{x})) = 0.
\end{cases}  
\end{equation}
where $\boldsymbol{x}$ denotes given initial conditions and the state variables to be solved are parameterized by the GNN $\boldsymbol{\phi}^{\boldsymbol{\theta}}({\boldsymbol x})$. $f_{PQV}$, $f_{DC}^{eq}$ and $f_{DC}^{Con}$ correspond to equality constraints of ($\ref{eq_1}$-$\ref{deqn_ex3}$), ($\ref{eq_4}$-$\ref{eq_6}$) and ($\ref{eq_7}$), respectively. \textcolor{black}{To model the process of tuning firing/extinction angle with rectifier/inverter transformers, (11) is given.} Here we only instantiate the control mode 1. For brevity, we use $f_{DC}$ to unifiedly denote $f_{DC}^{eq}$, $f_{DC}^{Con}$, and $f_{DC}^{K}$, and subscript $k$ is omitted.

\vspace{-1em}
\begin{subequations}
\label{eq_11} 
\begin{equation}
    \label{eq_11a}
    f_{DC}^{Ang}  (\boldsymbol{x}, \! \mathcal{\phi}^{\boldsymbol{\theta}} \! (  \boldsymbol{x}  ) )  \triangleq \text{ReLU}( \alpha^{min} -  \alpha  )
\end{equation}
\vspace{-3ex}
\begin{equation}
    \label{eq_11b}
    f\!_{DC}^{K}  \left(\boldsymbol{x}, \mathcal{\phi}^{\boldsymbol{\theta}} (  \boldsymbol{x} ) \right )  \triangleq \text{ReLU}(K^{re} -  K^{re, max}) + \text{ReLU} (K^{re, min} -  K^{re}) 
\end{equation}
\end{subequations}
\vspace{-3ex}

\textcolor{black}{The solution of (\ref{eq_10}) can then be approached by the Lagrangian duality method. Consider it that (\ref{eq_10}) is nonconvex and indeed hard to solve, we introduce the augmented Lagrangian method (ALM) to relief the solving difficulty \cite{r11-1}. ALM generally mitigates the strong convexity conditions, and can help us yield tolerable performance of the PG-GNN. By trial, we found the PG-GNN learned upon the ALM outperforms other competitive GNNs. Specifically, ALM rises the following learning scheme ($i \in \{ PQV,DC \}$).}
\begin{eqnarray}    
\label{eq_augL}
\textcolor{black}{ \mathcal{L}_{\rho}^{\boldsymbol{\theta}} \left( \boldsymbol{\lambda},\mathcal{\phi}^{\boldsymbol{\theta}}(\boldsymbol{x}) \right)}  &\textcolor{black}{=}& \textcolor{black}{{f}_{DC}^{Ang}\left(\mathcal{\phi}^{\boldsymbol{\theta}}(\boldsymbol{x}) \right) + \sum_ {i} \lambda_i {f}_i\left(\mathcal{\phi}^{\boldsymbol{\theta}}(\boldsymbol{x}) \right) }     \nonumber \\ 
\vspace{-2em}
&\textcolor{black}{+}& \textcolor{black}{ \frac{\rho}{2} 
\sum_{i} \left\Vert {f}_i\left(\mathcal{\phi}^{\boldsymbol{\theta}}(\boldsymbol{x}) \right) \right\Vert_2^2 }   
\end{eqnarray}
\vspace{-1.5em}
\begin{subequations}
    \label{ascent}
    \begin{equation}    \label{ascento}
    \textcolor{black}{\boldsymbol{\theta}^{k+1} :=\arg\min{\boldsymbol{\theta}} \mathcal{L}^{\boldsymbol{\theta}}_{\rho}\left( \boldsymbol{\lambda}^k,\mathcal{\phi}^{\boldsymbol{\theta}^{k}}(\boldsymbol{x}) \right)}
    \end{equation}
    \vspace{-1.2em}
    \begin{equation}
        \label{ascentmul}
        \textcolor{black}{\lambda_i^{k+1} :=\lambda_i^{k} + \rho \left | {f}_i \left( \mathcal{\phi}^{\boldsymbol{\theta}^{k+1}}(\boldsymbol{x}) \right) \right | }
    \end{equation}
\end{subequations}
\textcolor{black}{ where $\rho > 0$. (\ref{ascento}) updates $\boldsymbol{\theta}$ by minimizing (12). By fixing $\boldsymbol{\theta}$, (\ref{ascentmul}) then updates dual variables, which will be forwarded to perform (13a) over again. (13) continues until the setting steps are hit or (12) remains stable. Notably, (13a) can be realized via gradient descent:}

\vspace{-1.5em}
\begin{eqnarray}
    \label{eq_14}
    \boldsymbol{\theta} \leftarrow \boldsymbol{\theta} + \textcolor{black} { \frac{\partial \mathcal{L}^{\boldsymbol\theta}(\boldsymbol{\lambda},\mathcal{\phi}^{\boldsymbol{\theta}}(\boldsymbol{x}))}{\partial \boldsymbol{\theta}}} = \boldsymbol{\theta} + \textcolor{black}{\frac{\partial\mathcal{L}^{\boldsymbol\theta}(\boldsymbol{\lambda},\mathcal{\phi}^{\boldsymbol{\theta}}(\boldsymbol{x}))}{\partial \mathcal{\phi}^{\boldsymbol{\theta}}(\boldsymbol{x})} \frac{\partial \mathcal{\phi}^{\boldsymbol{\theta}}(\boldsymbol{x})} {\partial \boldsymbol\theta}} 
\end{eqnarray}

\begin{figure}[!t]
\setlength{\abovecaptionskip}{-0.2cm}  
\setlength{\belowcaptionskip}{-0.2cm} 
\centering
\includegraphics[width=3.5in]{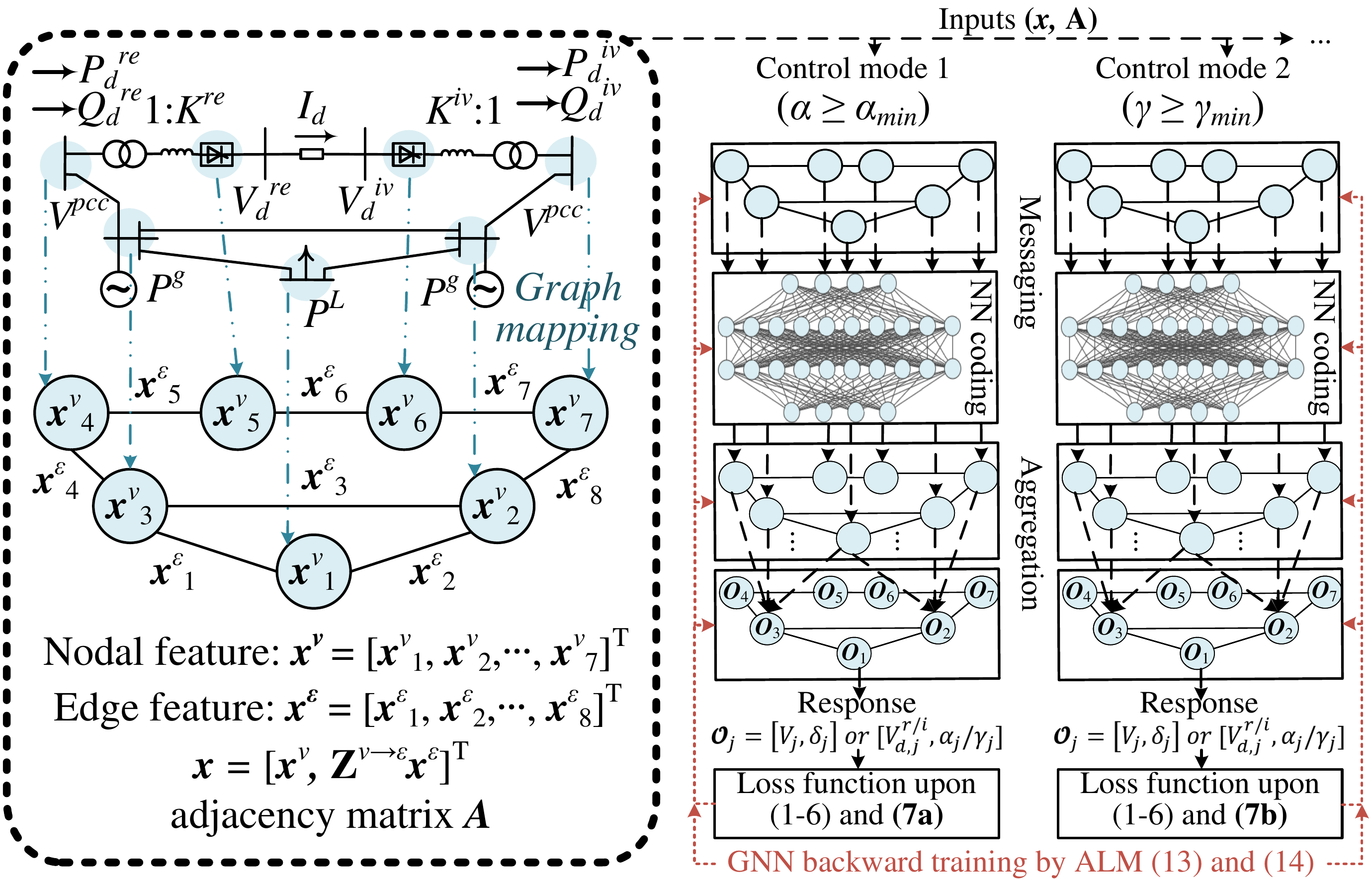}
\caption{The framework of the proposed method. The left subplot provides a toy example for describing our GNN structure, where a 5-bus and 5-branch system with 1 HVDC link is used. The GNN models the AC system via 5 nodes and 5 edges. Notably, to contain the information of the DC system, extra 2 nodes and 3 edges are involved in our GNN model. Finally, the GNN with $5+2 \times 1$ nodes and $5+3 \times 1$ edges is built}
\vspace{-1ex}
\label{fig1}
\vspace{-2ex}
\end{figure}

\textcolor{black}{Attentively, the GNN follows given DC control mode to render power flow solutions. We thus conduct multiple PG-GNNs to master DC control mode switching. Particularly, every control mode is traversed to train its specific PG-GNN. Then, all PG-GNNs compute power flow, and the one complies firing/extinction angle constraints will make the final decision. The implementation workflow is concluded in Fig. \ref{fig1}.}

\vspace{-0.8em}
\section{Case Study}
\subsection{Experimental settings}
The modified IEEE 30-bus system is conducted as the benchmark \cite{r7}, where a DC line is added between bus 2 and bus 4. Renewable energies are integrated on bus 6, 12, 10, 15 and 27, where the rated powers are 60MW, 70MW, 55MW, 65MW, and 65MW, respectively. A hydropower unit with rated capacity of 126MW is integrated at bus 2. The renewable penetration reaches 39.62$ \%$ of the installed capacity. Based on the historical hydropower, renewable energy and load profiles from a region \textcolor{black}{in southwest China}, 50,000 groups of data on renewable generation and load are gathered, with 75$\%$ severing as training data and the remaining as testing data.

\vspace{-1em}
\subsection{ Performance evaluation and comparative study }
\textcolor{black}{8} methods are introduced for comparative studies.

1) M0: The model-based method proposed in \cite{r3a}.

2) M1: A traditional data-driven baseline. The imitation learning based on multi-layer fully-connected NN \cite{r5}.

3) \textcolor{black}{M2: A physics-informed supervised learning \cite{r12}.} 

4) M3: The improved GNN-based imitation learning \cite{r7}. 

5) \textcolor{black}{M4: A GNN that takes the weighted sum of all AC/DC constraint violations as learning loss, where all penalty weights are subjectively selected. Such loss function refers to \cite{r10}.} 

6) M5: PG-GNN using graph convolutional network, where only one order neighbor information is aggregated.

7) M6: The proposed PG-GNN excluding edge feature $\boldsymbol{x}^{\mathcal{E}}$. 

8) \textcolor{black}{M7: A PG-GNN that mimics NR solver \cite{r11}.}

To justify the ability of the PG-GNN to master physical laws, \textcolor{black}{Tab. \ref{tab2} manifests constraint violation conditions}. From \textcolor{black}{Tab. \ref{tab2}}, our PG-GNN significantly beats other data-driven alternatives and performs nearly on par with the model-based approach. Three findings can be concluded: 1) on comparisons among M1-M3, and ours, shallowly imitating experience from data can cause severe violation of physics. 2) M1 and M3 shows the necessity of introducing intact topology in learning power flow patterns. 3) high-order neighbor information enable deeper learning on grid physics.
\textcolor{black}{Tab. \ref{tab2} also shows the elapsed time statistics of all applied methods. Observe that, M1-M6 beat M7, and are one order of magnitude faster than M0. This proves the efficiency merit of one-time feedforward computations. Notably, M1-M3 can be too idealistic to deploy in practice, since supervised learning demands copious time to prepare label data and enhance robustness, which is non-conductive to self-evolving in scenarios never seen before.}

\definecolor{MineShaft}{rgb}{0.2,0.2,0.2}
\begin{table}
\centering
\caption{Outcomes of constraint violation ($\times \rm 1e-5/ p.u.$) and elapsed time ($\rm{ms}$) by the applied methods ($mean \pm var.$)}
\vspace{-0.1cm}
\label{tab2} 
\begin{tblr}{c|c|c|c|c} Method & $\Vert\it f_{PQV} \Vert_{\text{1}}$ & $\Vert\it f_{DC}^{eq} \Vert_{\text{1}}$ & $\Vert\it f_{DC}^{Con} \Vert_{\text{1}}$ & Elapsed time \\ \hline M0 & $0 \pm 0$ & $0 \pm 0$ & $0 \pm 0$ & $126 \pm 18$ \\ M1 & $291 \pm 2325$ & $50 \pm 1562$ & $72 \pm 538$ & $12 \pm 10$ \\ \textcolor{black}{M2} & \textcolor{black}{$255 \pm 1092$} & \textcolor{black}{$41 \pm 489$} & \textcolor{black}{$42 \pm 407$} & \textcolor{black}{$13 \pm 10$} \\ M3 & $321 \pm 514$ & $35 \pm 302$ & $48 \pm 398$ & $12 \pm 9$ \\ \textcolor{black}{M4} & \textcolor{black}{$288 \pm 80$} & \textcolor{black}{$48 \pm 80$} & \textcolor{black}{$46 \pm 91$} & \textcolor{black}{$12 \pm 8$} \\ M5 & $208 \pm 472$ & $21 \pm 262$ & $17 \pm 65$ & $10 \pm 8$ \\ M6 & $110 \pm 108$ & $13 \pm 51$ & $17 \pm 32$ & $11 \pm 8$ \\ \textcolor{black}{M7} & \textcolor{black}{$79 \pm 68$} & \textcolor{black}{$1 \pm 50$} & \textcolor{black}{$1 \pm 35$} & \textcolor{black}{$48 \pm 30$} \\ Ours & $11 \pm 58$ & $1 \pm 45$ & $1 \pm 30$ & $12 \pm 10$ \end{tblr}
\vspace{-0.15cm}
\end{table}

\begin{figure}[!t]
\vspace{-0.2cm}  
\setlength{\abovecaptionskip}{-0.1cm}   
\setlength{\belowcaptionskip}{-0.6cm}   
\centering
\includegraphics[width=3.5 in]{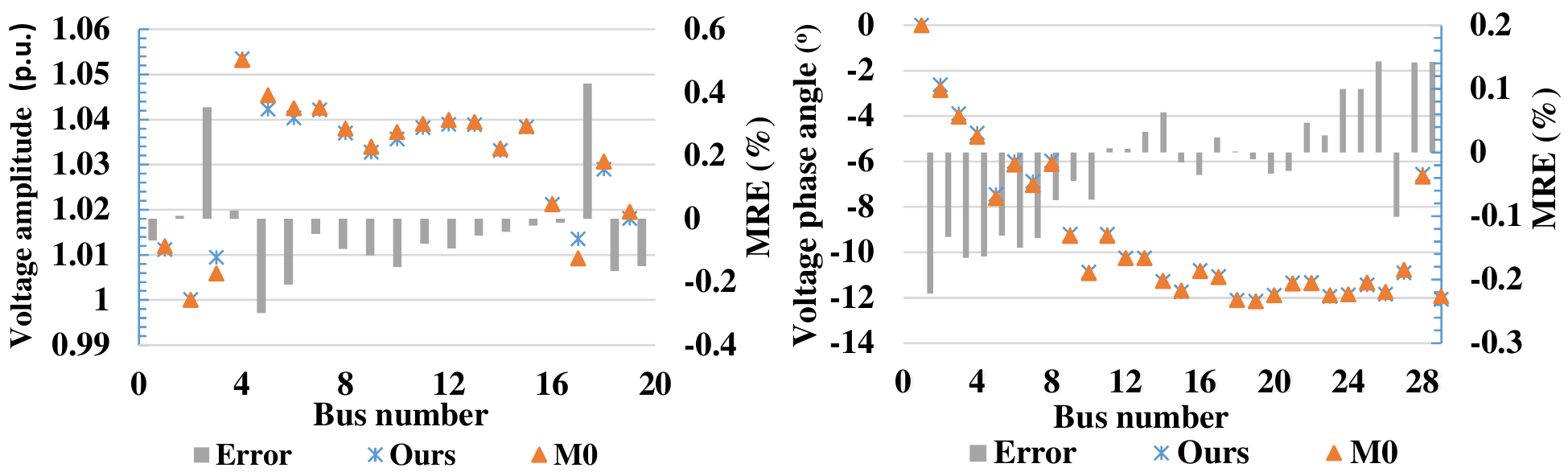}
\caption{The voltage amplitude and phase angle errors.}
\label{fig3}
\vspace{-0.2cm}
\end{figure}

 To clarify the performance of our PG-GNN,  \textcolor{black}{the mean relative error (MRE) of the bus voltage} by the PG-GNN against the baseline M0 are shown in Fig. \ref{fig3}. Fig. \ref{fig3} reveals that voltage amplitude error and phase angle error are less than $0.5  \% $ and $0.2 \%$, respectively. These statistics highlight the precision of our method.

\begin{figure}[!t]
\vspace{-0.2cm}  
\setlength{\abovecaptionskip}{-0.1cm}   
\setlength{\belowcaptionskip}{-2cm}   
\centering
\includegraphics[width=3.5 in]{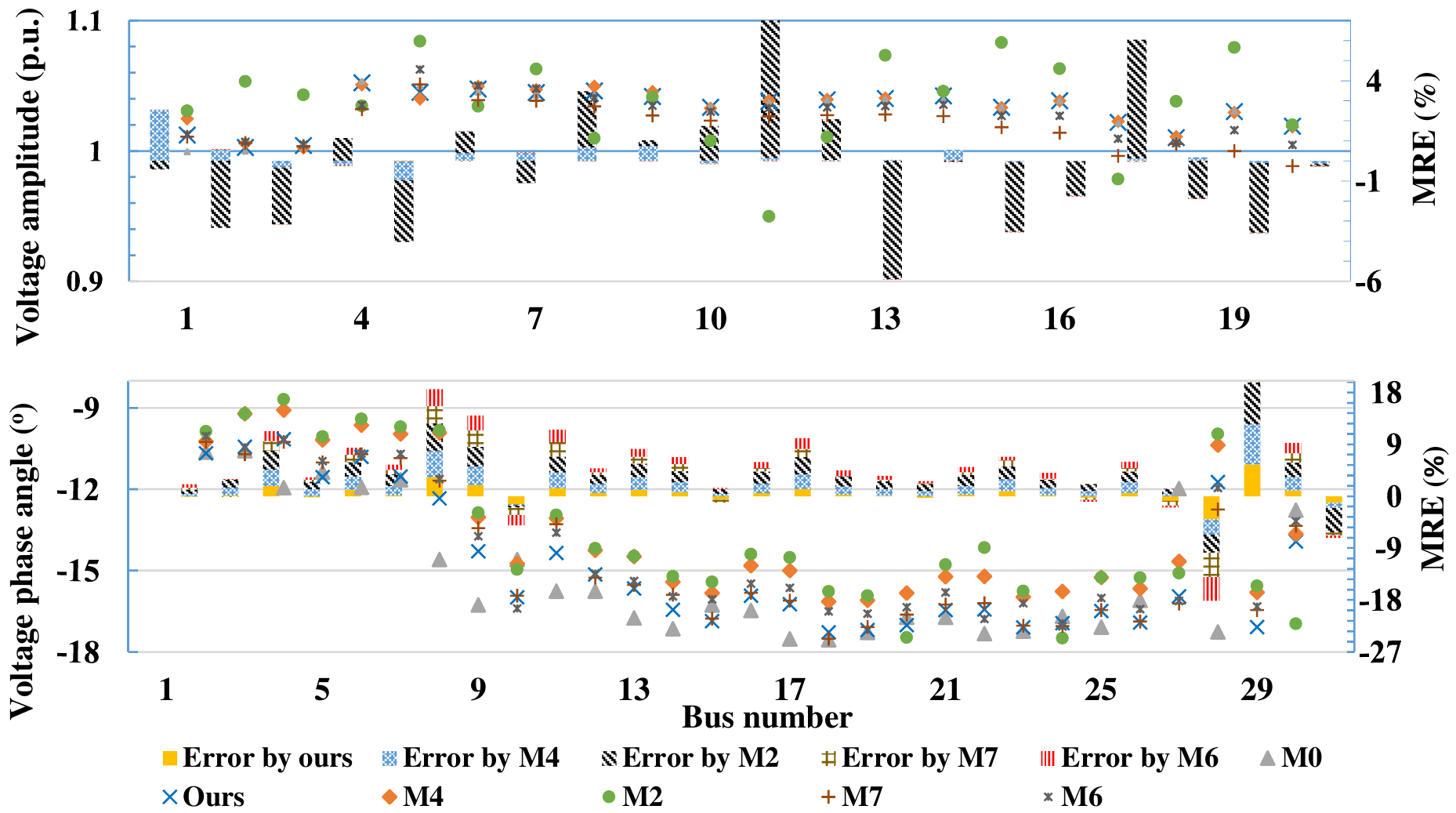}
\caption{The voltage amplitude and phase angle errors under branch switch-off.}
\label{fig5}
\vspace{-3ex}
\end{figure}

To discuss the topology adaptability, competitive trials among M2, M4, M6, M7 and ours are executed under branch 1-3 tripping. Outcomes are provided in Fig. \ref{fig5}. You can see that the PG-GNN is highly adaptive to topology change, and its performance can be enhanced by involving edge features.

\vspace{-0.8em}
\section{conclusion}
\textcolor{black}{A physics-guided graph neural networks (PG-GNN) is proposed for AC/DC power flow solution. A build-in graph model of AC and DC components is firstly designed. It then provides parameterized Lagrangian duality of AC/DC power flow problem to embed physics in PG-GNN. An augmented Lagrangian method (ALM)-based training strategy is proposed to enable PG-GNN to better fulfill nonconvex patterns. Decision framework is finally built upon multi-PG-GNN such that DC control mode switching can be manipulated. Case study on the IEEE 30-bus system justifies that the proposed method hits the best trade-off between decision efficiency and accuracy compared to classical model-based and 7 advanced data-driven methods. This work proved the necessity of physics embedding for reliable AC/DC power flow solution, and showed the flexibility of multiple PG-GNNs regarding model switching.}
\vspace{-1ex}
\bibliographystyle{IEEEtran}
\bibliography{paper}

\end{document}